\documentclass[twocolumn,aps,prl,showpacs,superscriptaddress,
amssymb]{revtex4}
\usepackage{color}
\usepackage{graphicx}% Include figure files
\usepackage{dcolumn}% Align table columns on decimal point
\usepackage{bm}% bold math
\usepackage{textcomp, pifont}
\begin{document}

%\preprint{APS/123-QED}

\title{Superconductivity with broken time reversal symmetry in ion irradiated Ba$_{0.27}$K$_{0.73}$Fe$_2$As$_2$ single crystals}
\author{V.\ Grinenko} \email{v.grinenko@ifw-dresden.de}
\affiliation{Institute for Solid State Physics, TU Dresden, 01069 Dresden, Germany}
\affiliation{IFW Dresden, Helmholtzstrasse 20, 1069 Dresden, Germany} 
\author{P.\ Materne}
\affiliation{Institute for Solid State Physics, TU Dresden, 01069 Dresden, Germany}
\author{R. Sarkar}
\affiliation{Institute for Solid State Physics, TU Dresden, 01069 Dresden, Germany}
\author{H. Luetkens} 
\affiliation{Laboratory for Muon Spin Spectroscopy, PSI, CH-5232 Villigen PSI, Switzerland}
\author{K. Kihou} 
\affiliation{National Institute of Advanced Industrial Science and Technology (AIST), Tsukuba, Ibaraki 305-8568, Japan}
\author{C. H. Lee} 
\affiliation{National Institute of Advanced Industrial Science and Technology (AIST), Tsukuba, Ibaraki 305-8568, Japan}
\author{S. Akhmadaliev} 
\affiliation{Helmholtz-Zentrum Dresden-Rossendorf, 01314 Dresden, Germany}
\author{D. V.\ Efremov}
\affiliation{IFW Dresden, Helmholtzstrasse 20, 1069 Dresden, Germany}
\author{S.-L.\ Drechsler}
\affiliation{IFW Dresden, Helmholtzstrasse 20, 1069 Dresden, Germany}
\author{H.-H.Klauss} 
\affiliation{Institute for Solid State Physics, TU Dresden, 01069 Dresden, Germany}

\date{\today}% It is always \today, today,

\begin{abstract}

Over the last years a lot of theoretical and experimental efforts have been made  to find states with broken time reversal symmetry (BTRS) in multi-band superconductors. In particular, it was theoretically proposed that in the  Ba$_{1-x}$K$_{x}$Fe$_2$As$_2$ system either an $s+is$ or an $s+id$ BTRS state may exist at high doping levels in a narrow region of the phase diagram. Here we report the observation of an enhanced zero field muon spin relaxation rate below the superconducting transition temperature for a high quality crystalline sample with $x \approx$ 0.73. This indicates that indeed the time reversal symmetry is broken in superconducting Ba$_{1-x}$K$_{x}$Fe$_2$As$_2$ at this doping level.

\end{abstract}


%
\maketitle 

The possibility to change the order parameter symmetry by charge doping in Fe based superconductors (FBS) recently attracted considerable attention. At optimal doping level most of the experimental and theoretical studies suggest an $s_{\pm}$ superconducting (SC) order parameter symmetry caused by commensurate spin fluctuations between hole and electron Fermi pockets.\cite{Hirschfeld2011} However, for high doping levels the available results are controversial. In the Ba$_{1-x}$K$_{x}$Fe$_2$As$_2$ system close to $x = 1$ some of the experimental data were interpreted in favor of $s$-wave \cite{Okazaki2012, Watanabe2013, Hardy2013a, Terashima2014, Ota2014} others support $d$-wave superconductivity \cite{Hashimoto2010, Reid2012, Abdel-Hafiez2013, Kim2014,Grinenko2014}. According to theoretical model calculations,\cite{Graser2009, Platt2013, Maiti2011} these states are almost degenerate at high hole doping level. However, the predicted $s_{\pm}$ state close to $x = 1$ is qualitatively different from the $s_{\pm}$ at optimal doping. For the former the order parameter changes sign between hole pockets.\cite{Maiti2013,Marciani2013} Indeed, various experimental investigations indicate that an essential change of the SC and the normal sate properties occur at a K doping of $x \sim 0.7$. Close to this doping level the electron pockets disappear and only the hole pockets remain on the Fermi surface according to angle-resolved photoemission spectroscopy (ARPES) measurements.\cite{Xu2013} Inelastic neutron scattering measurements revealed that the relation between $T_{\rm c}$ and the incommensurability $\delta$ of the low-energy spin fluctuation spectra changes its behavior abruptly around this doping level.\cite{Lee2016} At the same time, the spin resonance energy in the SC state falls below $2\Delta$ between $x$ = 0.7 and 0.8, where $\Delta$ is the SC gap-amplitude. Specific heat investigations show that the behavior of the specific heat jump ($\Delta C$) at $T_{\rm c}$ versus $T_{\rm c}$ deviates from the universal Bud'ko-Ni-Canfield (BNC) scaling around $x$ = 0.7.\cite{Budko2013,Hardy2016} 
Additionally, the thermal conductivity at low temperatures changes its behavior from exponential to linear between $x = 0.7$\ -\ $0.8$ suggesting the presence of accidental line nodes on some Fermi surface pockets.\cite{Watanabe2013} 

It was shown theoretically that the evolution of the order parameter with K doping can occur through intermediate $s+is$ or $s+id$ SC states that possesses an arbitrary phase on different Fermi surface sheets.\cite{Maiti2012,Maiti2013,Marciani2013,Maiti2015,Lin2016,Stanev2010}  In both states the time-reversal symmetry is broken and in the presence of non-magnetic defects spontaneous currents may emerge.\cite{Maiti2015,Lin2016} An $s+id$ state leads to local currents at any impurities below $T_{\rm c}$, \cite{Lee2009} while an $s+is$ state only induces local currents around impurities which locally break the tetragonal symmetry of the lattice.\cite{Maiti2015,Lin2016} Previous zero field $\mu$SR investigations of polycrystalline Ba$_{1-x}$K$_{x}$Fe$_2$As$_2$ samples with doping levels $x =$ 0.5, 0.6, 0.7, 0.8, and 0.9 did not reveal any noticeable enhancement of the muon spin relaxation rate below $T_{\rm c}$.\cite{Mahyari2014} These results exclude an $s+id$ and possibly an $s+is$ state at these doping levels since in a real material some of the crystalline defects such as dislocations and grain surfaces break the $C_4$ symmetry of the lattice. However, a BTRS state could be overlooked in this study since it can occupy a very narrow region in the phase diagram. Moreover,  non-magnetic disorder can narrow or even eliminate the region with a BTRS state since the $s_{\pm}$ superconductivity is sensitive to non-magnetic interband impurity scattering.\cite{Schilling2016}  Therefore,  experiments with single crystals with disorder under control are necessary to identify the intrinsic phase diagram of the Ba$_{1-x}$K$_{x}$Fe$_2$As$_2$ system. 

In our experiments, we focused on the doping region close to $x = 0.7$ where the SC order parameter may change its symmetry according to various experimental observations. Additionally, we used a low fluence ion irradiation to add a small amount of the symmetry breaking defects in high quality single crystals. Therefore, we target to detect a BTRS SC state, irrespective of the order parameter symmetry, by its local magnetic fields reflecting the local currents around defects or impurities. We observed that the zero field muon spin relaxation rate is enhanced below the temperature $T^* \sim 10$ K which is lower than $T_{\rm c} \sim 13$ K for the sample with the doping level $x \approx 0.73$. However, this behavior is absent for the samples with $x \approx 0.70$. The temperature dependence of the relaxation rate for $x \approx 0.73$ is consistent with theoretical predictions for both $s+is$ or $s+id$ SC states, while its small value points to an $s+is$ rather than an $s+id$ state. 

$\mu$SR experiments on the stacks of the Ba$_{1-x}$K$_{x}$Fe$_2$As$_2$ single crystals with doping levels of $x = 0.70(2)$ and 0.73(2) were performed at the GPS instrument of the $\pi$M3 beamline at the Paul Scherrer Institute (PSI) in Villigen, Switzerland. Fully spin-polarized, positive muons with an energy of 4.2 MeV were implanted in the sample (parallel to the crystallographic $c$-axis) where they rapidly thermalize and stop at interstitial lattice
sites at a depth of the order of $100$\ $\mu$m depending on the sample density. Given that it is very difficult to grow thick enough Ba$_{1-x}$K$_{x}$Fe$_2$As$_2$ single crystals in this doping range suitable for $\mu$SR experiments, we used stacks of several plate like (10\ -\ 50\ $\mu$m thick) single crystals with a total thickness $\sim 200$\ $\mu$m and an area between 3x3 and 4x4 mm$^2$. To ensure that the muons stop in the sample, we used an Al degrader with a thickness of $d_{\rm Al} = 300$\ $\mu$m. This value of $d_{\rm Al}$ is sufficient to reduce the muon energy to a value that all muons stop in the sample but not in the Al degrader. The whole assembly was wrapped into a very thin 5 $\mu$m Al foil and attached by a thin polyester tape to the Cu sample holder having the form of a fork. The Cu fork is designed in a such way that the muons do not hit the Cu holder. Zero field (ZF) and transverse field (TF) measurements were performed for two muon polarization modes. In the transverse polarization mode the so called up-down positron counters were used. The muon spin polarization $P_\mu$ is at about $45^{\circ}$ with respect to the muon beam (pointing toward the upward counter) and sample $c$-axis.\cite{PSI_GPS} In the longitudinal polarization mode with $P_\mu$ anti-parallel to the muon momentum (parallel to the sample $c$-axis) the backward and forward counters were used. The data were analyzed
using the musrfit software package.\cite{Suter2012} 

\begin{figure}[b]
\includegraphics[width=0.5\textwidth]{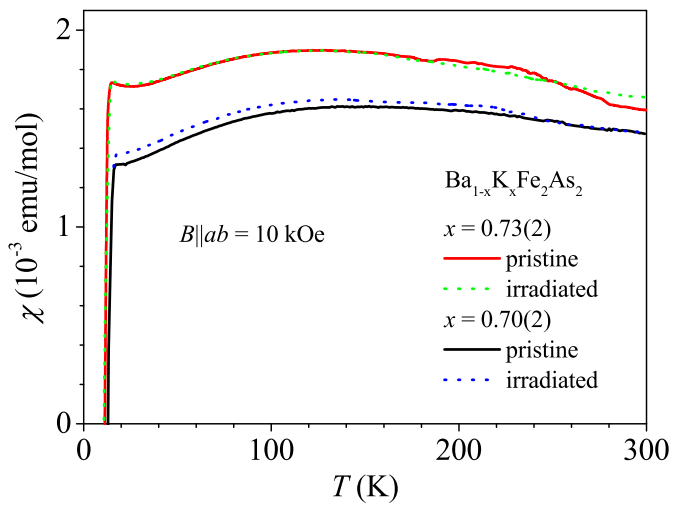}
\caption{(Color online) Temperature dependence of the molar
susceptibility $ \chi_m$ of the Ba$_{1-x}$K$_{x}$Fe$_2$As$_2$ single crystal stacks used in $\mu$SR experiments measured before (solid curves) and after (dashed curves) ion irradiation.} 
\label{Fig:1}
\end{figure}

The single crystals of two doping levels used in the experiments were selected from a single batch to ensure a similar sample quality. The $T_{\rm c}$ of each single crystal was checked by magnetization measurements using a commercial SQUID magnetometer from Quantum Design. The single crystals, which had $T_{\rm c}$ values within 1 K, were used to assemble the crystals stacks of each doping level. This $T_{\rm c}$ variation corresponds to 1 - 2 $\%$ of the doping difference between crystals in the samples used for the experiments. The susceptibility data of the stacks in the SC state are shown below (see Fig.\ \ref{Fig:3} right axes). The doping level was determined by x-ray diffraction using the known dependence of the $c$-axis lattice constant versus K doping (see Fig.\ 3 in Ref.\ \onlinecite{Kihou2016} and Fig.\ S1 in the Suppl.). The investigated single crystals have a high residual resistivity ratio of $RRR \approx 55$ measured before the irradiation (see Fig.\ S2 in the Suppl.) and a low amount of magnetic impurities indicated by the negligibly small upturn in the temperature dependence of the normal state magnetic susceptibility (Fig.\ \ref{Fig:1}). A low fluence $\sim 10^{10} {\rm cm}^2$ of I$^{+9}$ ions with an energy $\sim$ 50\ MeV with an angle of $60^{\circ}$ with respect to the single crystal $ab$-plane  was used to irradiate the samples. Each thin single crystal was irradiated from both sides to ensure that a large sample volume contains irradiation defects. Such an ion beam produces columnar defects (see Fig.\ 1 in Ref. \onlinecite{Fang2012}), which, in our case, are non-symmetrical in the $ab$-plane. These defects locally break the tetragonal $C_4$ symmetry of the lattice and, as predicted, should induce currents for both $s+is$ and $s+id$ symmetries. As one can see in Fig.\ \ref{Fig:3}, the irradiation did not affect $T_{\rm c}$ since the distance between defects of about 100 nm is much larger than the SC coherence length $\xi_{ab} = \Phi_0/2\pi H_{\rm c2} \sim 5$ nm and it is comparable with the electron mean free path (see Suppl.), where $H_{\rm c2} \sim 150$ kG for the field applied along crystallographic $c$ - axis.\cite{Liu2014} Therefore, it is safe to assume that the symmetry of the SC order parameter is not affected by the irradiation. Also, the irradiation did not affect the normal state magnetization as shown in Fig.\ \ref{Fig:1}. However, the diamagnetic response below $T_{\rm c}$ in the field cooled branch of the low field magnetic susceptibility is noticeably suppressed after the ion irradiation (Fig.\ \ref{Fig:3}). This suppression is most likely caused by the trapping of the vortices by irradiation defects.

In the $\mu$SR experiments we focused on the irradiated samples. Prior to the ZF-$\mu$SR measurements we always performed TF measurements of each sample in the normal and SC state to estimate the amount and the relaxation rate of the background signal. The real part of the Fast Fourier Transform (FFT) is shown in Figs.\ \ref{Fig:2}a\ -\ \ref{Fig:2}c. The data analysis indicates that in all cases there is a contribution of a non-relaxing background signal attributed to muons stopping outside of the sample and presumably in the cryostat walls. In the case of the sample with $x$ = 0.73 (with mass $m_{\rm s1} = 14.6$ mg) the background signal contribution was around 13 - 14 \% depending on the polarization mode and the sample mounting. However, in the case of the smaller sample ($m_{\rm s2} = 10.5$ mg) with $x$ = 0.70 a larger background signal ($\approx 30$ \%) was determined. The obtained values of the background signal were used for analysis of the ZF data. 

\begin{figure}[b]
\includegraphics[width=0.5\textwidth]{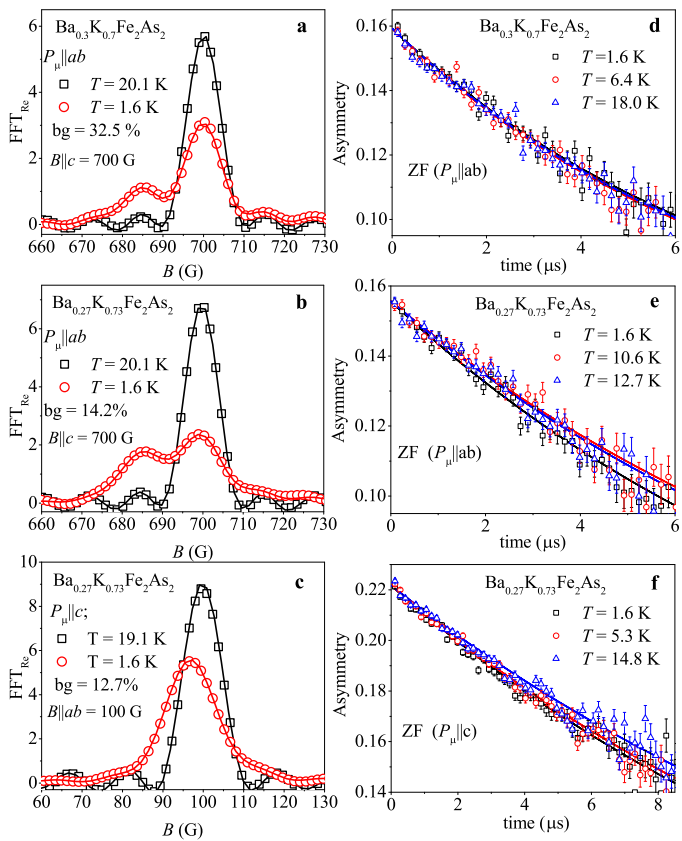}
\caption{(Color online) Real part of the FFT of the transverse-field $\mu$SR time spectra obtained above and below $T_{\rm c}$ for irradiated Ba$_{1-x}$K$_{x}$Fe$_2$As$_2$ samples: (a) $x$ = 0.70 , (b) $x$ = 0.73 measured in the transverse polarization mode, and (c) $x$ = 0.73 measured in the longitudinal polarization mode. The solid curves are fits to extract the background contribution to the $\mu$SR data. Representative ZF-$\mu$SR asymmetry time spectra of Ba$_{1-x}$K$_{x}$Fe$_2$As$_2$ for (d) $x$ = 0.70, (e) $x$ = 0.73 measured in the transverse polarization mode, and (f) $x$ = 0.73 measured in the longitudinal polarization mode. The solid curves are fits to Eq.\ \ref{Eq1}, described in the text.} \label{Fig:2}
\end{figure}

ZF measurements were performed under true zero field conditions (with the compensated earth-field) established after TF measurements at $T \sim 20$ K above $T_{\rm c}$. Then the samples were cooled down to the base temperature of around 1.6 K. Each series of the measurements was performed with ascending temperature steps with a temperature stability better than 0.1 K. In the transversal polarization mode 3 million counts per detector were measured at each temperature point and 15 million were measured in the longitudinal polarization mode. ZF-$\mu$SR asymmetry spectra of Ba$_{1-x}$K$_{x}$Fe$_2$As$_2$ at different temperatures are shown in Figs. \ref{Fig:2}d\ -\ \ref{Fig:2}f. In the case of the sample with $x =$ 0.70 the asymmetry spectra are essentially temperature independent. For the sample with $x =$ 0.73 we observe a decrease of the relaxation rate with the increase of the temperature (see also Fig.\ \ref{Fig:3}). A similar behavior was observed for both polarization modes. To fit the asymmetry spectra the simplest possible model was used:
\begin{equation}
A(t) = A_{\rm s}(0){\rm exp}[-\lambda/t]+A_{\rm bg},\label{Eq1}
\end{equation} 
where $A_{\rm s}(0)$ is the initial sample asymmetry, $\lambda$ is the relaxation rate, and $A_{\rm bg}$ is the non-relaxing background asymmetry obtained from TF measurements. The results of the fit by Eq.\ \ref{Eq1} are shown by the solid lines in Figs. \ref{Fig:2}d\ -\ \ref{Fig:2}f. The relaxation rate of our samples ($\lambda \approx 0.1$\ ${\mu}s^{-1}$) for $P_{\mu} \parallel ab$ is similar to a total relaxation rate of the polycrystalline samples from Ref.\ \cite{Mahyari2014} and slightly lower for $P_{\mu} \parallel c$ with $\lambda \approx 0.05$\ ${\mu}s^{-1}$. However, in our case the relaxation shows an exponential behavior in contrast to the dominant Gaussian contribution in the polycrystalline samples attributed to a nuclear relaxation. Recently a weak exponential relaxation was observed also for optimally and slightly overdoped single crystals in Ref.\ \cite{Mallett2016}. A weaker relaxation rate for $P_{\mu} \parallel c$ of our crystals as compared to the polycrystalline samples cannot be naively attributed to an impurity contribution additional to the Gaussian relaxation due to randomly oriented nuclear moments. Therefore, the discrepancy between the relaxation behaviors of single crystals and polycrystalline samples is not clear so far and requires further investigations.

\begin{figure}[b]
\includegraphics[width=0.5\textwidth]{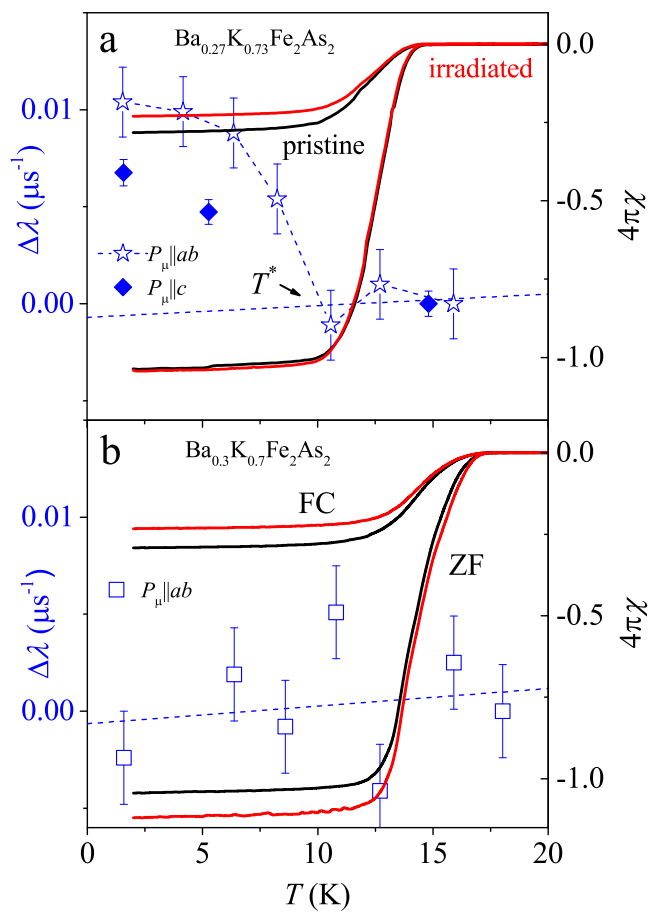}
\caption{(Color online) (Left axis) Temperature dependence of the relaxation rate $\Delta\lambda = \lambda(T) - \lambda_0$  for Ba$_{1-x}$K$_{x}$Fe$_2$As$_2$ samples: (a) $x$ = 0.73, where $\lambda_0(T = 16{\rm K}) = 0.085(2)$\ ${\mu}s^{-1}$ for $P_{\mu} \parallel ab$ and $0.054(1)$\ ${\mu}s^{-1}$ for $P_{\mu} \parallel c$, (b) $x$ = 0.70, where $\lambda_0(T = 18{\rm K}) = 0.131(3)$\ ${\mu}s^{-1}$ for $P_{\mu} \parallel ab$. The dashed curves are guide to the eyes. (Right axis) Temperature dependence of the volume susceptibility of the same samples before and after ion irradiation measured in a low magnetic field $B\parallel ab =$ 5 G applied after cooling in ZF, subsequent warming in the field and cooling again in the same field (FC).}
\label{Fig:3}
\end{figure}  

The obtained temperature dependence of the relaxation rate  $\Delta\lambda = \lambda(T) - \lambda_0$ for both samples together with the low field susceptibility data is shown in Fig.\ \ref{Fig:3}, where $\lambda_0$ is the relaxation rate above $T_{\rm c}$. $\Delta\lambda$ is enhanced at $ T^* \sim 10$ K $< T_{\rm c}$ for the sample with $x =$ 0.73 while it is nearly temperature independent for the sample with $x =$ 0.70. The $\Delta\lambda$ value and the sharp enhancement are consistent with the appearance of weak magnetic fields due to impurity induced currents in a BTRS SC state. Typically,  $\Delta\lambda$, associated with this state, varies between 0.005 and 0.05 $\mu {\rm sec}^{-1}$ for various superconductors.\cite{Heffner1990, Luke1998, Aoki2003, Hillier2009, Maisuradze2010, Shu2011, Hillier2012, Biswas2013, Singh2014, Zhang2015, Barker2015} 
In most of these superconductors a time reversal symmetry is broken already at $T_{\rm c}$, which can be explained within a single-band approach. The known exception is U$_{1-x}$Th$_x$Be$_{13}$  where a BTRS state appears at the second $T_{\rm c2} < T_{\rm c1}$.\cite{Heffner1990} So far the nature of the second transition in U$_{1-x}$Th$_x$Be$_{13}$ is under debate.\cite{Kromer1998,Sonier2000} There are arguments in favor of a magnetic origin of this transition based on specific heat and thermal expansion measurements.\cite{Kromer1998} In contrast to U$_{1-x}$Th$_x$Be$_{13}$, for our samples we didn't observe additional anomalies in the specific heat around $T^*$ (see Fig.\ S3 in the Suppl.). The lack of a clear anomaly imposes constrains on possible types of a BTRS state in Ba$_{1-x}$K$_{x}$Fe$_2$As$_2$. In particular, an $s+id$ state proposed in Refs. \onlinecite{Lee2009, Platt2013} is unlikely. In this state both $s$ and $d$ order parameters are present with a relative phase shift $\pm \pi/2$. These order parameters are weakly coupled due to different symmetries and therefore, may appear at different temperatures. Hence, for an $s+id$ state one expects an anomaly in the specific heat at $T^*$ similar to a transition at $T_{\rm c}$. In contrast, the transition between an $s_{\pm}$ to an $s+is$ state may not result in a noticeable anomaly at $T^*$ depending on the system parameters. In this case, gradual phase evolution of the gaps on different Fermi pockets can lead to a broad anomaly below $T_{\rm c}$, which can be hardly distinguished from a specific heat of a multi-band $s$-wave superconductor. Also, according to calculations given in Ref. \onlinecite{Lin2016} the local internal fields associated with an $s+is$ state are expected to be in the range of $(10^{-8} - 10^{-4})H_{\rm c2} \sim 0.001 - 10$ G depending on the defect potential, while for an $s+id$ state the internal fields are about $10^3 - 10^5$ times stronger. The experimental value of the average internal field $\Delta\lambda/\gamma_{\mu} \sim 0.1$ Oe is too small for an $s+id$ state even considering the diluted irradiation defects, only, where $\gamma_{\mu} = 0.085 {\mu}s^{-1} G^{-1}$ is the muon gyromagnetic ratio. Therefore, our data favor $s+is$ symmetry of the SC order parameter in Ba$_{1-x}$K$_{x}$Fe$_2$As$_2$ system at $x \sim 0.7$. This conclusion is consistent with a very recent theoretical study.\cite{Boeker2017} However,  for an unambiguous experimental discrimination between the $s+is$ and $s+id$ SC state further investigations are necessary such as, systematic studies of the effect of ion irradiation on the value of the relaxation rate in the BTRS state. Additionally, the measurements of the local field distribution around non-magnetic defects using a magnetic force microscope or a SC quantum interference device may help to determine directly the symmetry of the order parameter in the BTRS state.\cite{Lin2016}     

Finally we conclude that ZF-$\mu$SR measurements of ion irradiated moderately hole overdoped Ba$_{1-x}$K$_{x}$Fe$_2$As$_2$ single crystals revealed a possible broken time reversal symmetry in the SC state. This BTRS state forms below the bulk SC transition temperature $T_{\rm c}$ in qualitative agreement with theoretical predictions. Our as well as published data \cite{Mahyari2014} suggest that the possible BTRS states occupy a narrow region in between $x =$ 0.7 and 0.8 in the phase diagram of Ba$_{1-x}$K$_{x}$Fe$_2$As$_2$.

This work was supported by the DFG through grant DFG (GR 4667/1-1) and within the research training group GRK 1621. R.S. and H.H.K. are thankful to DFG for the financial assistance through the SFB 1143 for the project C02. This work was, also, supported by a Grant-in-Aid for Scientific Research B (No.\ 24340090) from the Japan Society for the Promotion of Science. S.-L.D. and D.E. thank the VW-fundation for partial support. This work was partially performed at Swiss Muon Source (S$\mu$S), PSI, Villigen. We acknowledge the fruitful discussion with A.\ Amato, P.\ Chekhonin, K.\ Iida, S.\ Ishida, S.\ Kamusella, and Y.\ Yerin.

\maketitle 

\appendix
\renewcommand\thefigure{\thesection S\arabic{figure}}    
\section{Supplementary materials}
\setcounter{figure}{0}

\maketitle

In this Supplementary, we provide additional 
x-ray,  electrical resistivity and specific heat data for the B$_{1-x}$K$_x$Fe$_2$As$_2$ single crystals used in our $\mu$SR experiments.

\begin{figure}[t]
\includegraphics[width=0.5\textwidth] {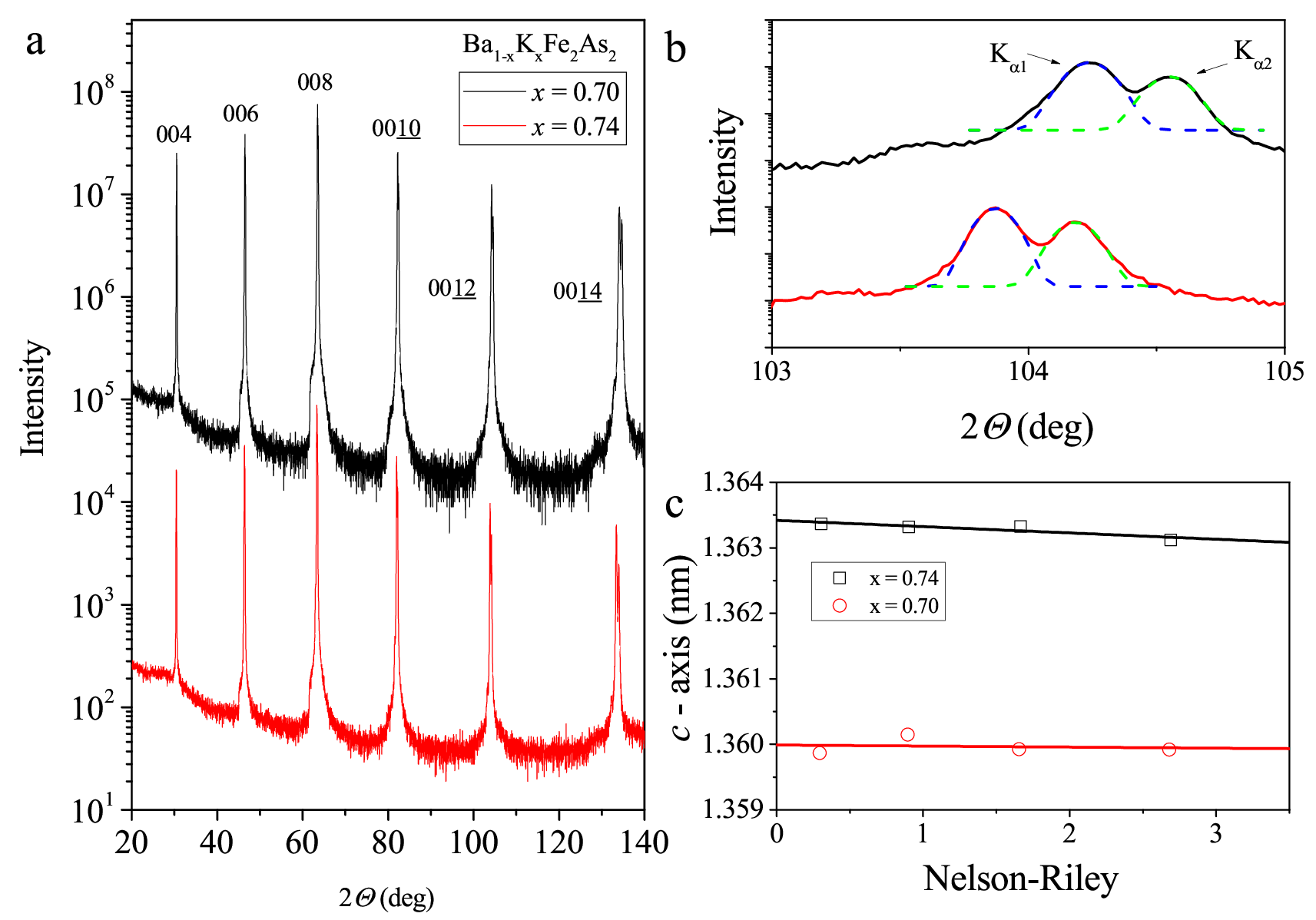}
\caption{(a) The $\Theta - 2\Theta$ x-ray diffraction patterns of two pristine (non-irradiated) Ba$_{1-x}$K$_x$Fe$_2$As$_2$ single crystals from the $\mu$SR samples $x =$ 0.70(2) and $x =$ 0.73(2). (b) The (012) reflections of the same samples. The Co K$_{\alpha 1}$ and K$_{\alpha 2}$ are clearly resolved indicating the high quality of the crystals. (c) The $c$-axis lattice parameter of the individual peaks versus the Nelson-Riley function. The crossing of the linear fit with the $y$-axis gives the $c$-axis value of the crystal.}
\label{XRD}
\end{figure}

\vspace{0.2cm}
The K doping level of the single crystals was calculated using the relation between the $c$-axis lattice parameter and the K doping obtained in the previous studies.\cite{Kihou2016S} The $c$-axis lattice parameter of the Ba$_{1-x}$K$_x$Fe$_2$As$_2$ single crystals were calculated from x-ray $\Theta - 2\Theta$ diffractograms using the Nelson-Riley function to reduce possible experimental errors as shown in Fig.\ \ref{XRD}. Any noticeable effect of the ion irradiation on the x-ray spectra was not observed.  
\vspace{0.2cm}

\begin{figure}[b]
\includegraphics[width=18pc,clip]{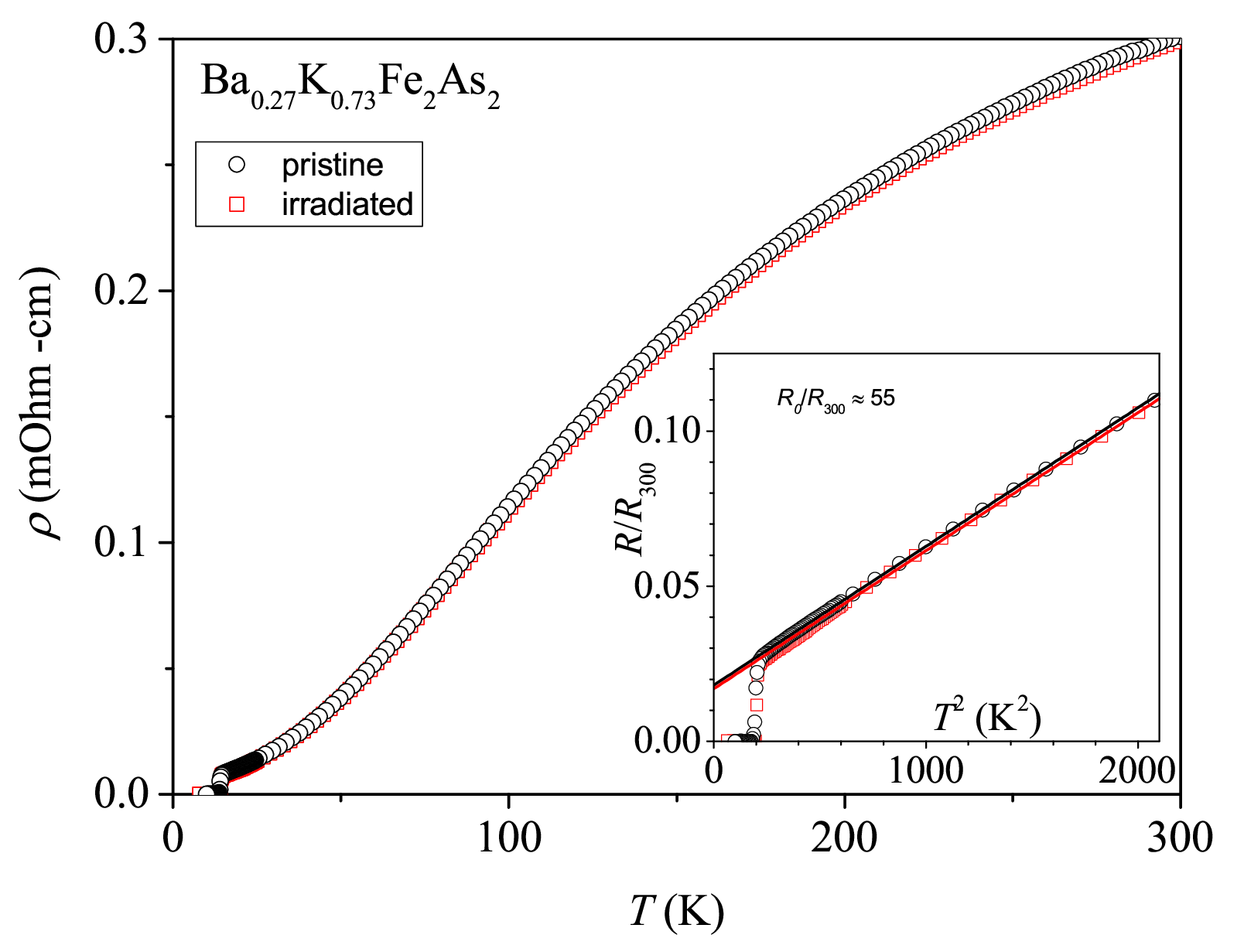}
\caption{Temperature dependence of the resistivity of the pristine (non-irradiated) and irradiated sample of Ba$_{1-x}$K$_x$Fe$_2$As$_2$ single crystal with the similar doping level $x = 0.73(1)$  (the pristine crystal is the same as in Fig.\ref{XRD}) in zero magnetic field. The resistivity versus $T^2$ is shown in the inset.}
\label{res}
\end{figure}

\vspace{0.2cm}
The resistivity data of our non-irradiated and irradiated  Ba$_{1-x}$K$_x$Fe$_2$As$_2$ single crystals with a similar doping level $x = 0.73(1)$ are shown in Fig.\ \ref{res}. The pristine crystal has a high residual resistivity ratio value of $RRR = \rho_{300}/\rho_{0} \approx 55$ with a $T^2$-Fermi-liquid behavior below 50 K indicating a low amount of impurities and disorder, where $\rho_{300}$ is the resistivity at $T$ = 300 K and $\rho_{0}$ is the extrapolated residual resistivity as shown in the inset of Fig.\ \ref{res}. The residual resistivity value $\rho_{0} \approx 5 \mu \Omega$ corresponds to the electron mean free path $l = m^*v_{F}/ne^2\rho_{0} \sim 10 - 100$ nm estimated using the reported values of the effective mass $m^* \sim 5 - 10 m_e$, the carrier concentration $n \sim 10^{-27} m^3$, and the Fermi velocity $v_F \sim 4\cdotp 10^4 m/s$ for the hole overdoped Ba$_{1-x}$K$_x$Fe$_2$As$_2$ compound.\cite{Johnston2010S, Terashima2010S} As one can see in Fig.\ \ref{res} the irradiation does not affect noticeably the $RRR$ value since the distance between defect $\gtrsim l$, justifying that the crystals remain in a clean limit after the irradiation.   
 
\vspace{0.2cm}

\newpage

\begin{figure}[t]
\includegraphics[width=15pc,clip]{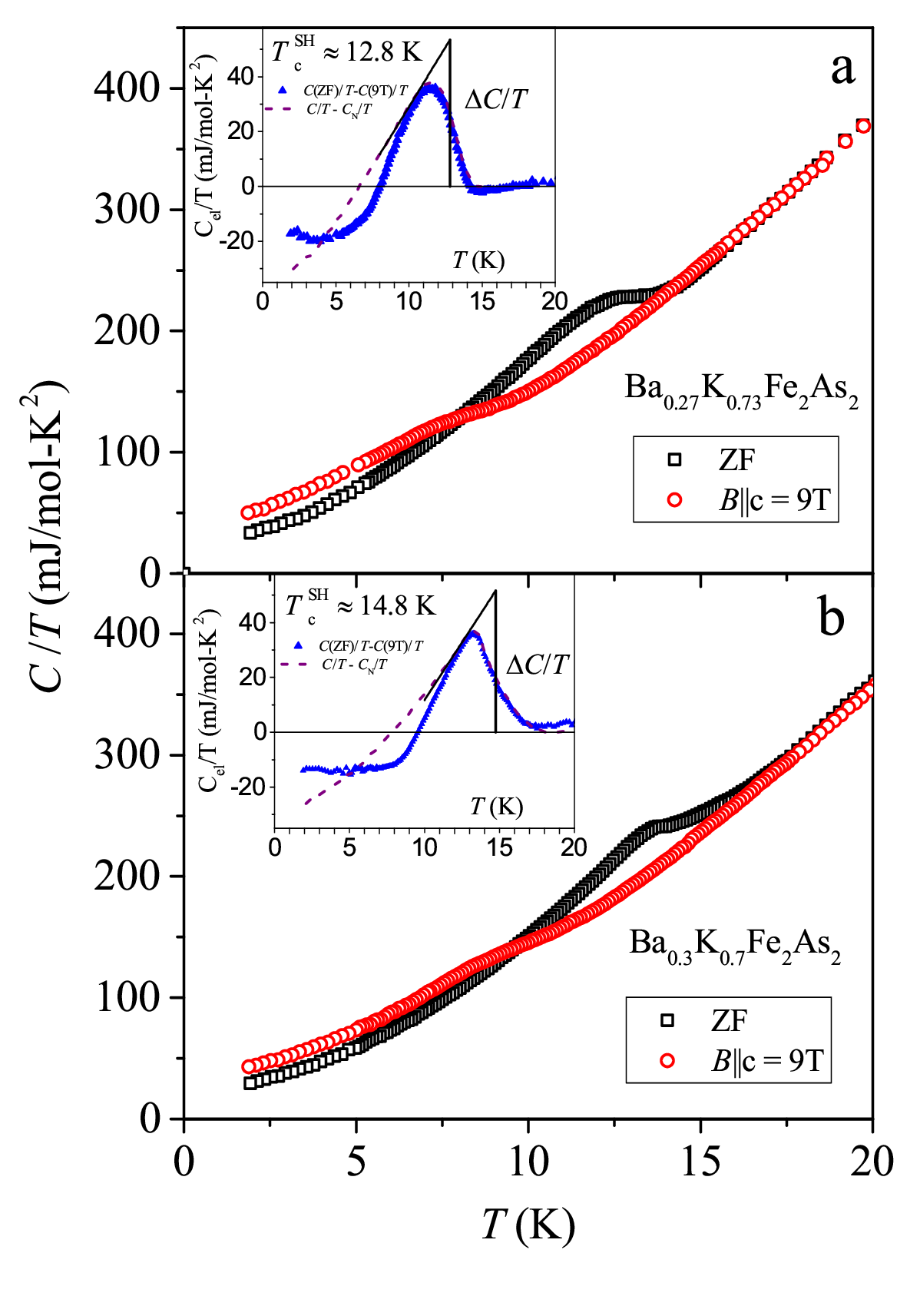}
\caption{Temperature dependence of the specific heat of the irradiated Ba$_{1-x}$K$_{x}$Fe$_2$As$_2$ single crystals stacks used in $\mu$SR experiments, (a) $x = 0.73(2)$ and  (b) $x = 0.70(2)$. Open symbols - measurements in zero and an applied magnetic field of 9 T along the $c$-axis of the crystals. Insets show the difference between the ZF and the 90 kG data - closed symbols and the difference between the specific heat in the superconducting and extrapolated normal state - doted lines.}
\label{SH}
\end{figure}

\vspace{0.2cm}
The specific heat of the irradiated Ba$_{1-x}$K$_{x}$Fe$_2$As$_2$
single crystals stacks used in our $\mu$SR experiments are shown in Fig.\ \ref{SH}. The samples exhibit a clear superconducting anomaly in agreement with the susceptibility measurements (main text). The transition width $\Delta T_{\rm c}^{\rm SH} \sim$ 2 - 3 K is related to $T_{\rm c}$ variation of the single crystals within a stack and to the transition width of the individual crystals of about 1 - 2 K for these doping levels. No additional anomalies were observed below $T_{\rm c}$. The transition temperature estimated from the entropy constriction is $T_{\rm c}^{\rm SH} \approx $ 12.8 K  for $x =$ 0.73 and 14.8 K $x =$ 0.70.  The estimated electronic specific heat in the normal state (Sommerfeld coefficient) are about $\gamma_{\rm el} \approx 62(3)$ mJ/mol-K$^2$ with the specific heat jump $\Delta C/T_{\rm c}^{\rm SH} \approx 53(1)$ mJ/mol-K$^2$ at $T_{\rm c}^{\rm SH}$ for $x =$ 0.73 and $\gamma_{\rm el} \approx 53(3)$ mJ/mol-K$^2$ with the specific heat jump $\Delta C/T_{\rm c}^{\rm SH} \approx 52(1)$ mJ/mol-K$^2$ at $T_{\rm c}^{\rm SH}$ for $x =$ 0.70. These values are in agreement with those of previous measurements.\cite{Budko2013S,Hardy2016S}

\vspace{0.2cm}

\end{thebibliography}

\end{document}